\documentclass[fleqn,12pt,twoside]{article}
\usepackage[headings]{espcrc1}

\usepackage{graphicx}
\usepackage{epsfig}

\usepackage{color}

\newcommand{\AmS}{{\protect\the\textfont2
  A\kern-.1667em\lower.5ex\hbox{M}\kern-.125emS}}

\title{Parton cascade and coalescence}

\author{Denes Molnar\address[OSU]{Physics Department, Ohio State University\\
        191 West Woodruff Avenue, Columbus, OH 43210, USA}%
        \address[PU]{Department of Physics, Purdue University\\
        525 Northwestern Avenue, West Lafayette, IN 47907, USA}%
        \address[BNL]{RIKEN/BNL Research Center\\
        Brookhaven National Laboratory, Upton, NY 11973-5000, USA}}

\newcommand{\vp}{{\vec p}}

\newcommand{\vx}{{\vec x}} 
\newcommand{\vv}{{\vec v}}

\newcommand{\gton}{\mathrel{\lower.9ex \hbox{$\stackrel{\displaystyle 
>}{\sim}$}}}  
\newcommand{\lton}{\mathrel{\lower.9ex \hbox{$\stackrel{\displaystyle 
<}{\sim}$}}}

\newcommand{\be}{\begin{equation}} 
\newcommand{\ee}{\end{equation}}  
\newcommand{\ben}{\begin{enumerate}}  
\newcommand{\een}{\end{enumerate}} 
\newcommand{\bit}{\begin{itemize}}  
\newcommand{\eit}{\end{itemize}} 
\newcommand{\bc}{\begin{center}}  
\newcommand{\ec}{\end{center}} 
\newcommand{\bea}{\begin{eqnarray}}  
\newcommand{\eea}{\end{eqnarray}} 

\begin{document}

\maketitle

\begin{abstract}
This is a review of the parton cascade approach and its implications 
on parton 
coalescence at RHIC.
\end{abstract}

\section{Introduction}
Recent data from $Au+Au$ reactions at $\sqrt{s_{NN}}=130$ and $200$ 
GeV from RHIC show striking difference between baryons and mesons
in the intermediate transverse
momentum region $2 < p_\perp < 5$ GeV.
First, elliptic flow scales with 
constituent quark number%
\cite{QM2005exp,STARboth,PHENIXv2scaling,STARkstar,STARxiomega}.
The flow per constituent quark as a function of transverse momentum
per constituent quark, $v_2^{hadron}(p_T/n)/n$ 
($n=2$ for mesons, $3$ for baryons),
is a universal function for all hadron species 
(within experimental uncertainties).
Second, nuclear suppression is weaker ($R_{AA}$ is larger) 
for baryons than for mesons%
\cite{QM2005exp,STARboth,PHENIXnoBsupp,STARnoBsupp,STARkstar,STARphi,%
PHENIXphi}.
The most promising mechanism proposed to explain both phenomena is
parton coalescence%
\cite{ALCORMICOR,Voloshincoal,HwaYang,texbudMtoB,dukeCoal,coalv2,%
charmcoal,dyncoal}.

In the coalescence model,
mesons form from a quark and antiquark, 
while baryons from three quarks or three antiquarks
(contributions from higher Fock states are small \cite{DukeFOCK}).
Most versions of the model are based on variants of the 
simple ``coalescence formula''~%
\cite{Dover,Scheibl,texbudMtoB,dukeCoal,charmcoal}
\be
\frac{dN_{had}(\vp)}{d^3p} \! =\! g_M \! 
\int\! \prod_{i}\left[ d^3 x_i d^3 p_i f_{i}(\vp_i,\vx_i)\right] \, 
W_{had}(\{\Delta \vx_{ij}\},\{\Delta \vp_{ij}\})\,
\delta^3(\vp{-}\sum_{i}\vp_i)
\label{coaleq}
\ee
that gives the hadron spectra in terms of the 
constituent phasespace distributions $f_i$ on a 3D spacetime hypersurface
and the Wigner-transform of hadron wave functions $W$.
(Arbitrary 3D hypersurfaces\cite{Dover,Scheibl}
and quark correlations\cite{DukeCORR} are straightforward to accommodate.)
This simple approach can  
reproduce the particle spectra at RHIC quite well,
with coalescence hadronization from a thermalized quark-antiquark 
plasma\cite{texbudMtoB,dukeCoal}
and an additive fragmentation contribution of quenched high-$p_T$ 
jets\cite{partonEloss}.
The scaling of elliptic flow with constituent quark
number was also explained in this framework\cite{coalv2,charmcoal}.

Nevertheless, several important questions are still open.
At low $p_T$, 
the coalescence formula violates\cite{coalv2,charmcoal} unitarity.
The yield in a given coalescence channel scales quadratically/cubically
with constituent number,
moreover, the same constituent contributes to several channels
(including fragmentation in certain schemes).
Energy conservation is only approximate because of the on-shell treatment
and neglect of binding energies.
Also, though
the final entropy is comparable to the initial one (due to hadron decays) 
\cite{TexasEntropy},
it is problematic that entropy does decrease temporarily during the 
coalescence process.

Most importantly, the simple parametrizations assumed for the quark 
phase space distributions are inconsistent with dynamical models, 
such as parton cascades 
or hydrodynamics.
Spatial inhomogeneities and dynamical phasespace correlations
distort coalescence predictions in a crucial way.
In particular, the observed quark number scaling becomes highly 
nontrivial~\cite{v2cvsh},
and a large baryon/meson ratio enhancement is difficult to 
achieve\cite{dyncoal}.

Below I review and illustrate the difficulties, using one of the main 
dynamical approaches, covariant parton transport theory.

\section{Covariant parton transport theory}

Covariant parton transport theory is the incoherent, particle 
(short-wavelength) limit of quantum-chromodynamics.
Its main advantage is that it is applicable out of
equilibrium and models freezeout self-consistently.  However, it
cannot describe phase transitions (without coupling to classical
fields).

One of the simplest but nonlinear
forms of Lorentz-covariant Boltzmann transport 
theory involves on-shell phase space densities $\{f_i(x,\vp)\}$
that evolve with elastic $2\to 2$ \cite{ZPC,ZPCv2,nonequil,v2,dissipv2} 
and {\em inelastic} $2\to 2$ \cite{charmv2} rates as
\be
p_1^\mu \partial_\mu f_{1,i} = S_i(x, \vp_1) +
\frac{1}{16\pi^2}\sum\limits_{jk\ell} 
\int\limits_2\!\!\!\!
\int\limits_3\!\!\!\!
\int\limits_4\!\!
\left(
f_{3,k} f_{4,\ell} \frac{g_i g_j}{g_k g_\ell} - f_{1,i} f_{2,j}
\right)
\left|\bar{\cal M}_{12\to 34}^{ij\to k\ell}\right|^2 
\delta^4(p_1+p_2-p_3-p_4)
 \ .
\label{Boltzmann_eq}
\ee
$|\bar{\cal M}|^2$ is the polarization averaged scattering matrix 
element squared,
the integrals are shorthands
for $\int\limits_a \equiv \int d^3 p_a / (2 E_a)$,
$g_i$ is the number of internal degrees of freedom 
for species $i$,
while $f_{a,i} \equiv f_i(x, \vp_a)$. 
The source terms 
$\{S_i(x,\vp)\}$ specify the initial conditions.

Eq.~(\ref{Boltzmann_eq}) could in principle be extended for bosons
and/or for inelastic processes, such as $gg\leftrightarrow ggg$~\cite{inel,XG}.
However, with the new nonlinearities these extensions introduce,
it is very challenging to maintain Lorentz covariance numerically
at opacities expected at RHIC.
Off-shell variants of parton transport also exist, 
such as the VNI/b model\cite{VNIb}.

Based on the local mean free path
$\lambda(x,s) \equiv 1/[n(x)\sigma(s)]$, transport 
theory naturally interpolates between ideal hydrodynamics $(\lambda \to 0)$
and free streaming $(\lambda \to \infty)$.
The most relevant quantity that characterizes the whole evolution is
the transport opacity $\chi  \equiv \int dz \rho(z) \sigma_{tr} $ \cite{v2}, 
which is the average number of collisions 
per parton multiplied by the ratio of the transport and total cross sections 
(the latter is the efficiency of momentum transfer in a single 
collision).
In a near-equilibrium situation, these parameters can be related to transport 
coefficients, such as the shear viscosity or diffusion constants.
\begin{figure}[hbpt] 
\begin{center}
\epsfig{file=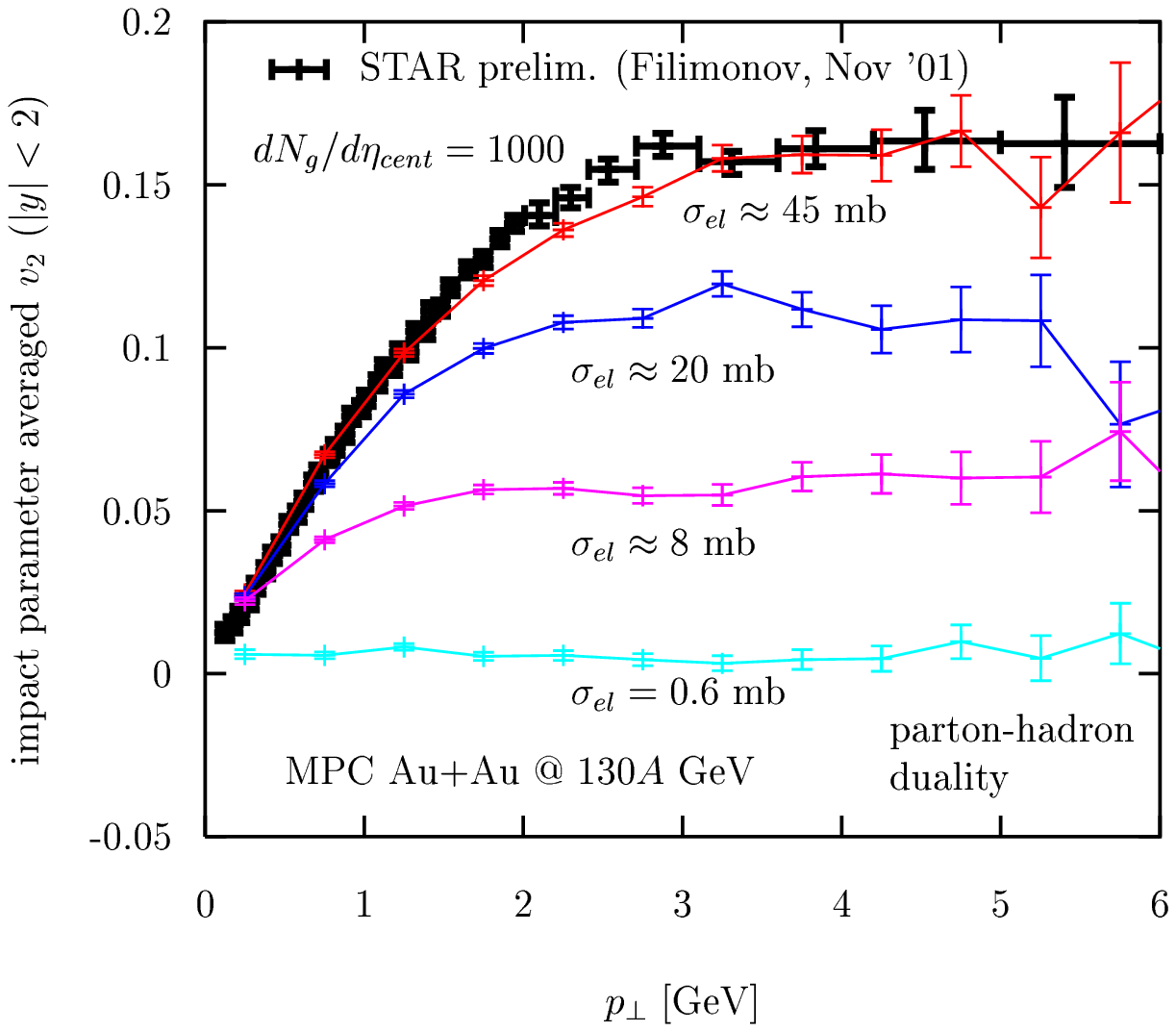,height=2.2in,width=2.5in,clip=5,angle=0}
\epsfig{file=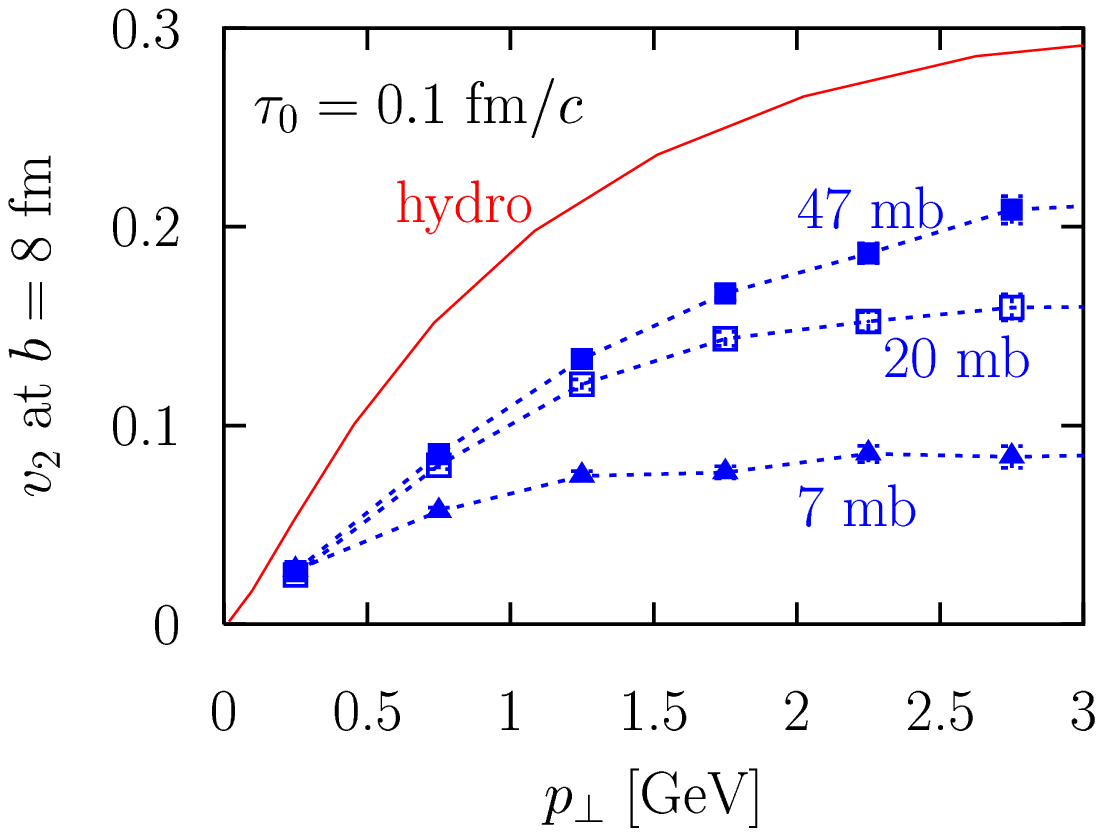,height=2.2in,width=2.3in,clip=5,%
angle=0}
\end{center}
\vspace*{-0.25cm}
\hskip 3.8cm a) \hskip 5.85cm b)
\vspace*{-0.4cm} 
\caption{\label{fig:1}
Results for $Au+Au$ at $\sqrt{s}=200A$ GeV at RHIC with $b=8$ fm from
MPC\cite{MPC}.
a) Transport opacity dependence of elliptic flow as a function of $p_T$ 
from covariant transport theory \cite{v2};
b) comparison of elliptic flow as a function of $p_T$ from covariant transport
theory and ideal hydrodynamics \cite{dissipv2}.
}
\end{figure}

Transport theory provides important information about the properties of 
the partonic medium created at RHIC.
The large elliptic flow observed%
\cite{STARv2charged,PHENIXv2charged,PHOBOSv2charged} indicates a 
strongly-interacting, opaque parton system\cite{v2}, 
with $dN_g/d\eta(b=0) \times \sigma_{gg} \approx 1000\times 45$ mb,
about 15 times above the perturbative estimate 
(see Fig. \ref{fig:1}a).
These conditions are still different from an ideal fluid
because dissipation reduces elliptic flow by $30-50$\% relative 
to the ideal fluid limit \cite{dissipv2}
(Fig. \ref{fig:1}b). At such high opacities,
the shear viscosity of the parton plasma $\eta \sim s \lambda T/5$ 
\cite{MiklosPawel85} (estimated from kinetic theory)
is indeed very small, close to the conjectured lower bound 
$\eta_{min} = s/(4\pi)$ \cite{etalimit} ($s$ is the entropy density). 
However, in heavy ion collisions,
the small viscosity is compensated by large gradients, 
resulting in significant dissipative effects.
Estimates based on viscous (Navier-Stokes) hydrodynamics also support this
conclusion\cite{Teaneyviscos}.

\begin{figure}[h] 
\parbox[b]{7cm}{
\epsfig{file=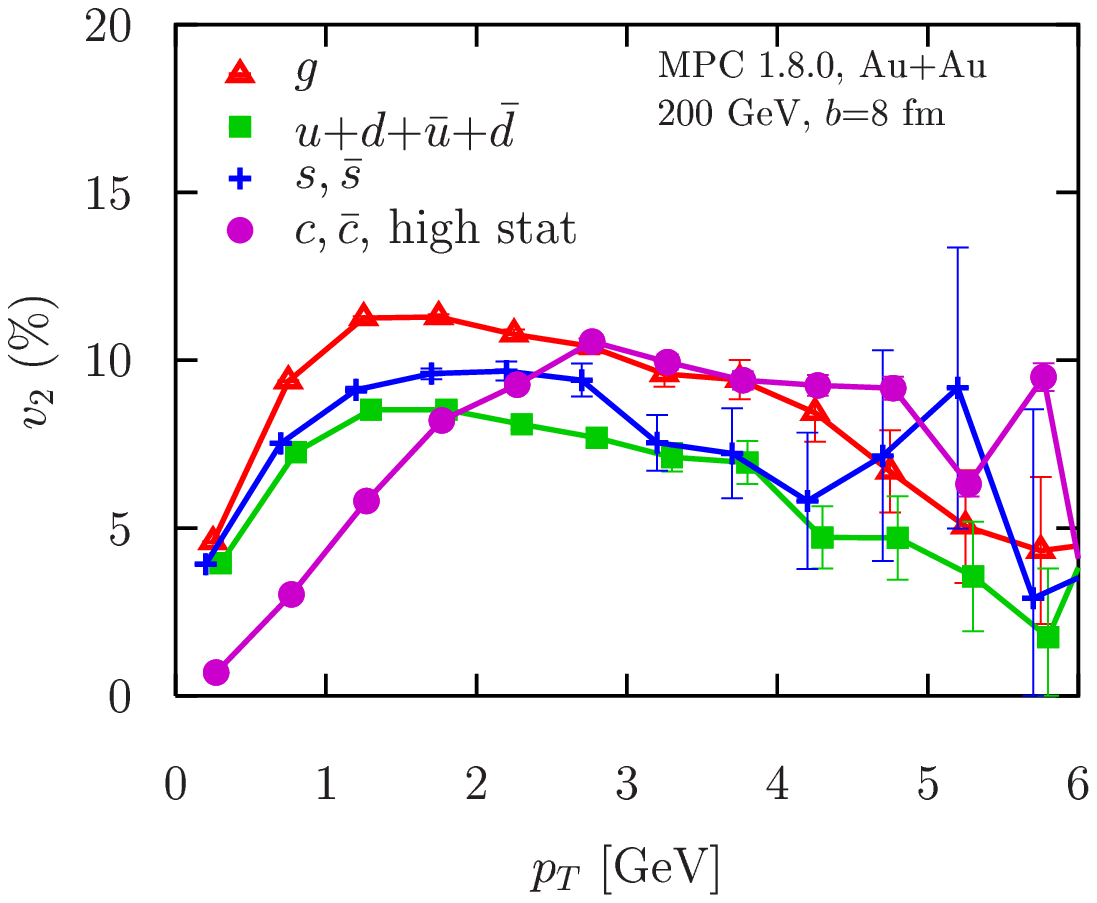,height=1.8in,width=2.5in,clip=5,angle=0} \\
\epsfig{file=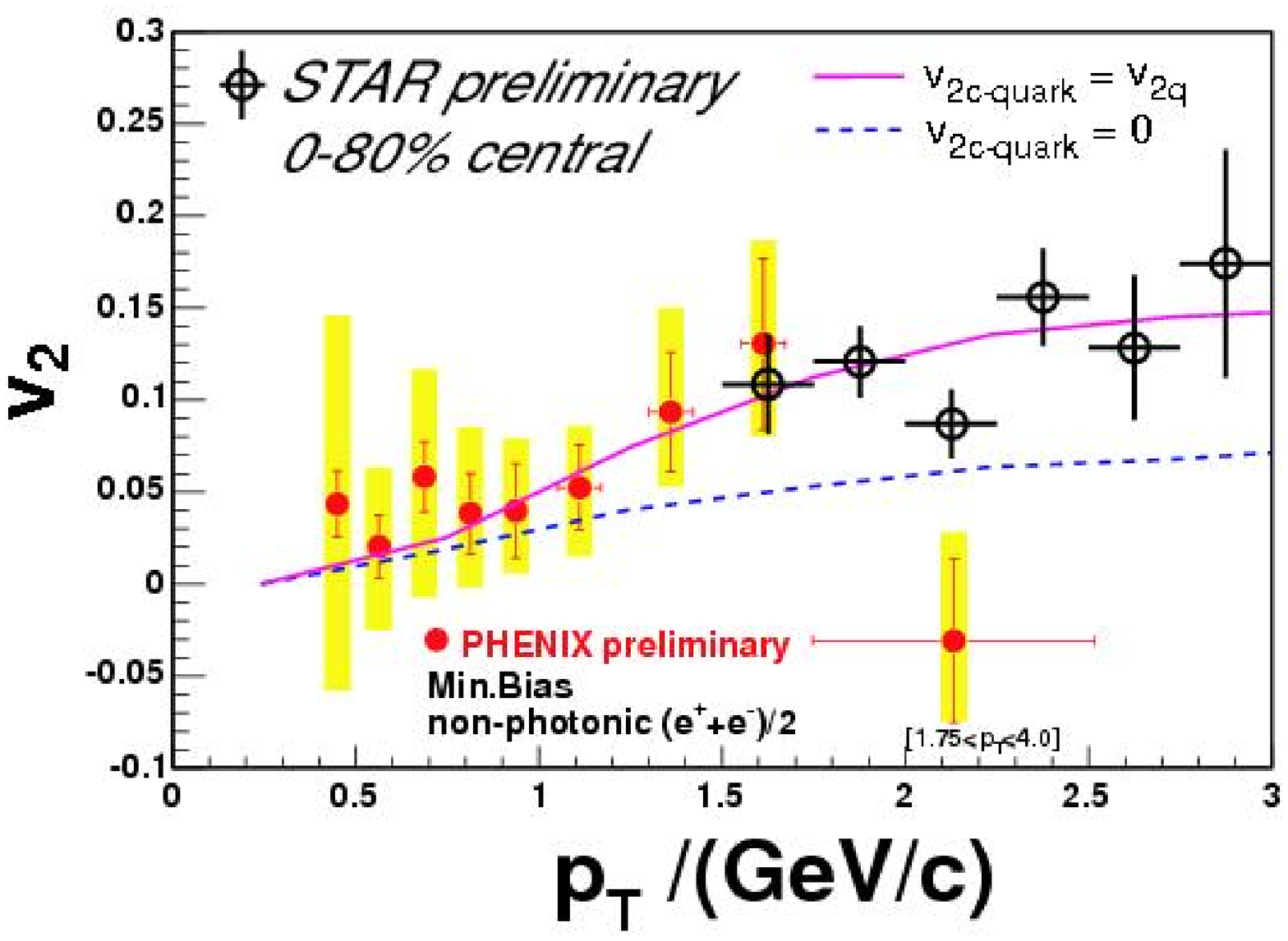,height=1.8in,width=2.3in,clip=5,angle=0}
}
\hskip -0cm
\parbox[b]{7.5cm}{
\epsfig{file=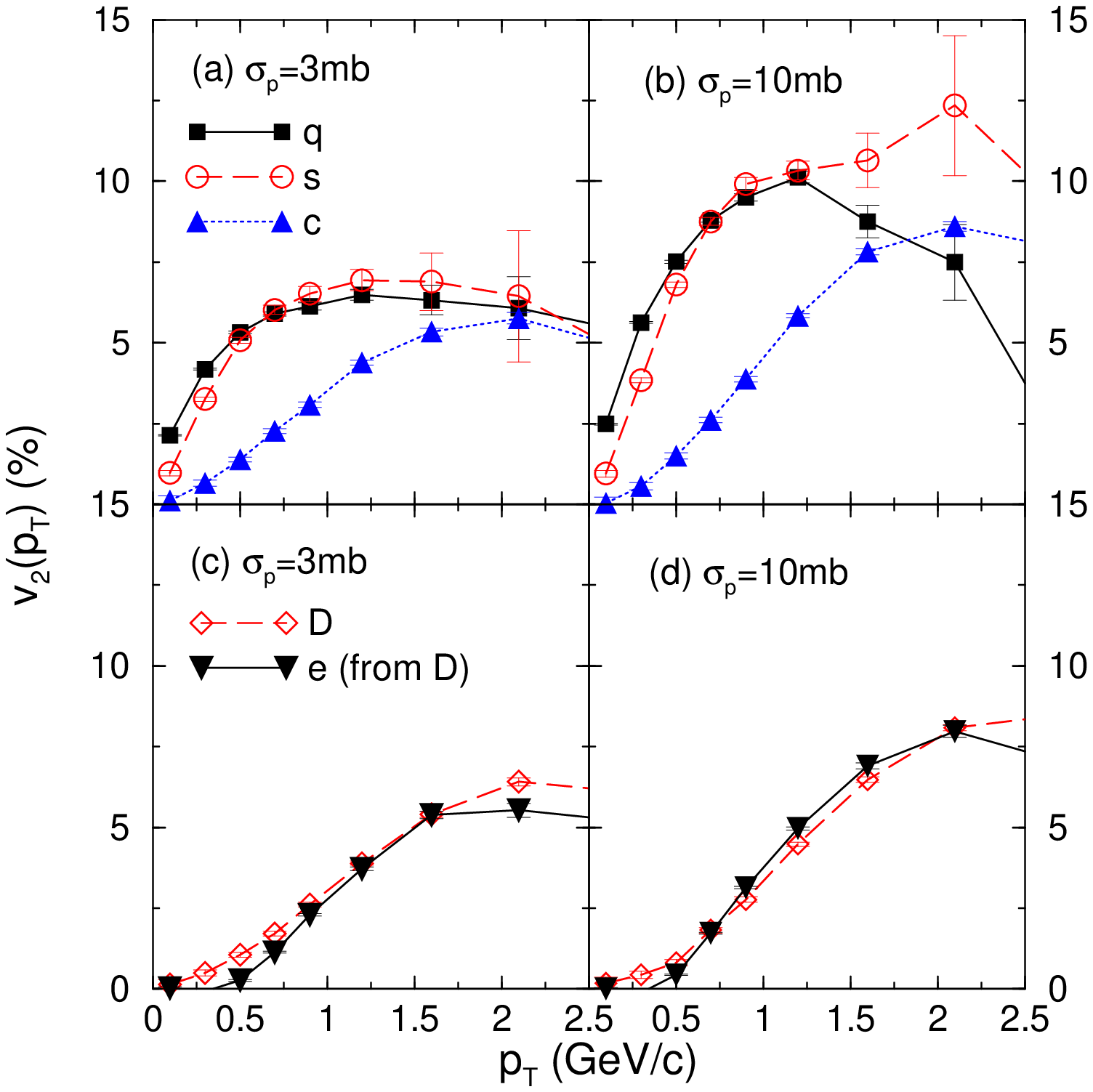,height=3in,width=3.45in,clip=5,angle=0}
}\\
\vspace*{-0.25cm}
\hskip 3.8cm a) \hskip 5.85cm b)
\vspace*{-0.4cm} 
\caption{ \label{fig:2}
a) Top panel: parton species dependence of elliptic flow in $Au+Au$ 
at $\sqrt{s_{NN}} = 200$ GeV at RHIC
as a function of $p_T$, for $b=8$ fm and $\approx 7$ times perturbative 
opacities, from the covariant MPC model \cite{charmv2};
bottom panel:
preliminary 
non-photonic electron $v_2(p_T)$ data from RHIC \cite{STARcharmv2}.
b) Top panels: parton $v_2(p_T)$ for minimum-bias $Au+Au$ collisions at 
$\sqrt{s_{NN}}=200$ GeV from 
the AMPT model; bottom panels: comparison of $D$ meson $v_2$
and electron $v_2$ from $D$ decays \cite{Bincharmv2}.
}
\end{figure}

Though even the large opacities at RHIC are insufficient to equilibrate 
heavy quarks,
a large charm quark elliptic flow 
is still expected above $p_T \sim 2-3$ GeV
(Fig.~\ref{fig:2})
from several studies 
(covariant transport predictions from MPC\cite{charmv2,MPC},
reinforced by calculations from AMPT\cite{Bincharmv2} 
and also a Fokker-Planck approach\cite{Teaneycharm}).
It is very exciting that 
preliminary indirect data from RHIC do indicate 
a sizeable charm elliptic 
flow\cite{STARcharmv2,PHENIXcharmv2}.
These data are on the elliptic flow of electrons from $D$ meson 
decays $D^{(*)}\to K \nu e$, which based on model studies follows closely 
the $D$ meson elliptic flow\cite{Texascharm,Bincharmv2} 
(see Fig.~\ref{fig:2}b).

At high $p_T$, covariant transport shares a lot of similarity with
parton energy loss models\cite{partonEloss},
except that $2\to 2$ transport gives {\em incoherent, elastic} energy loss.
However, at the high opacities at RHIC the dynamics is much richer.
Not only energy loss, but significant energy gain is also possible in multiple
collisions\cite{plasmapush}.
This ``plasma push'' process, a shadow of near-hydrodynamic
behavior at low $p_T$, could play 
a role even at $p_T \sim 10$ GeV,
and is the reason why elliptic flow from the transport decreases very 
slowly at high $p_T$.

\section{Dynamical effects and elliptic flow scaling from coalescence}

To illustrate the effect of the dynamics on elliptic flow scaling from
coalescence,
it is useful to first ignore
variations of phasespace distributions on length and momentum scales 
corresponding to a typical hadron ($\sim 1$ fm and $\sim 200$ MeV).
In this case\cite{v2cvsh}, 
$W \sim \delta^3(\Delta \vec x_{ij}) \delta^3(\Delta \vec p_{ij})$ and
the phasespace distributions of mesons and baryons
from coalescence ($\alpha\beta \to M$, $\alpha\beta\gamma \to B$) are
\bea
 f_B(x,\vp) &=& \frac{(2\pi)^3 g_B}{g_\alpha\, g_\beta\, g_\gamma} \,
                f_\alpha(x,\vp/3) \, f_\beta(x,\vp/3)
                f_\gamma(x,\vp/3)
\nonumber \\
 f_M(x,\vp) &=& \frac{(2\pi)^6 g_M}{g_\alpha\, g_\beta} \, 
                f_\alpha(x,\vp/2) \, f_\beta(x,\vp/2)  \ .
\label{coaleq2}
\eea
$g$ is the degeneracy of the particle (spin and color),
and I considered constituent quarks of (roughly) equal mass, which share the 
total hadron momentum (roughly) equally.
(In hadrons that also contain 
a heavy quark, e.g., $D$ mesons or the $\Lambda_c$,
the heavy quark carries most of the momentum~\cite{charmcoal}.)

Assume for simplicity that all quarks have {\em identical} phase space
distributions
\be
f_q(x,\vp_T,y=0) \equiv n_q(x,p_T)\left[1+\sum_{n=1}^{\infty} 
2 v_{n,q}(x,p_T) \cos (n\phi)\right] \ ,
\ee
where $n_q$ is the local density of quarks with transverse momentum 
{\em magnitude} $p_T$.
For small local anisotropies 
\be 
|v_{2,q}| \ll 1 \ , \qquad |v_{k,q} v_{l,q}| \ll |v_{2,q}| \ ,
\label{scalingcond}
\ee
the usual assumption to obtain elliptic flow scaling, 
the hadron elliptic flows from (\ref{coaleq2}) are
\bea
v_{2,M}(p_T) 
&\approx & \frac{2\, \langle n_q^2(x,p_T/2)\, v_{2,q}(x,p_T/2) \rangle_x}%
                {\langle n_q^2(x,p_T/2)\rangle_x} 
\nonumber\\
v_{2,B}(p_T) 
&\approx& \frac{3\, \langle n_q^3(x,p_T/3)\, v_{2,q}(x,p_T/3) \rangle_x}%
               {\langle n_q^3(x,p_T/3)\rangle_x} 
\label{v2scaling}
\eea
with $\langle A(x) \rangle_x \equiv \int d^3 x A(x)$ (spatial average). 
Though each small spatial region with its own local quark elliptic flow 
contributes a local hadron elliptic flow that scales with quark number,
$v_{2,had}(x,p_T) \approx n\, v_{2,q}(x,p_T/n)$,
quark number scaling does not follow {\em in general} for
the measured (spatially averaged) elliptic flow,
i.e., $v_{2,had}(p_T) \ne n\, v_{2,q}(p_T/n)$.
This is because constituent phase space densities appear nonlinearly,
raised to second and third power.

Constituent quark scaling arises only for special classes of 
quark phase space distributions. Two classes where 
the spatial dependence in (\ref{coaleq2}) cancels out trivially are 
i) when the local density is spatially uniform $n(x,p_T) = n(p_T)$ 
(as assumed in \cite{coalv2});
and ii) when the local flow
anisotropy is spatially uniform $v_2(x,p_T) = v_2(p_T)$ (the assumption in 
\cite{dukeCoal}).

\begin{figure}[hbpt] 
\begin{center}
\epsfig{file=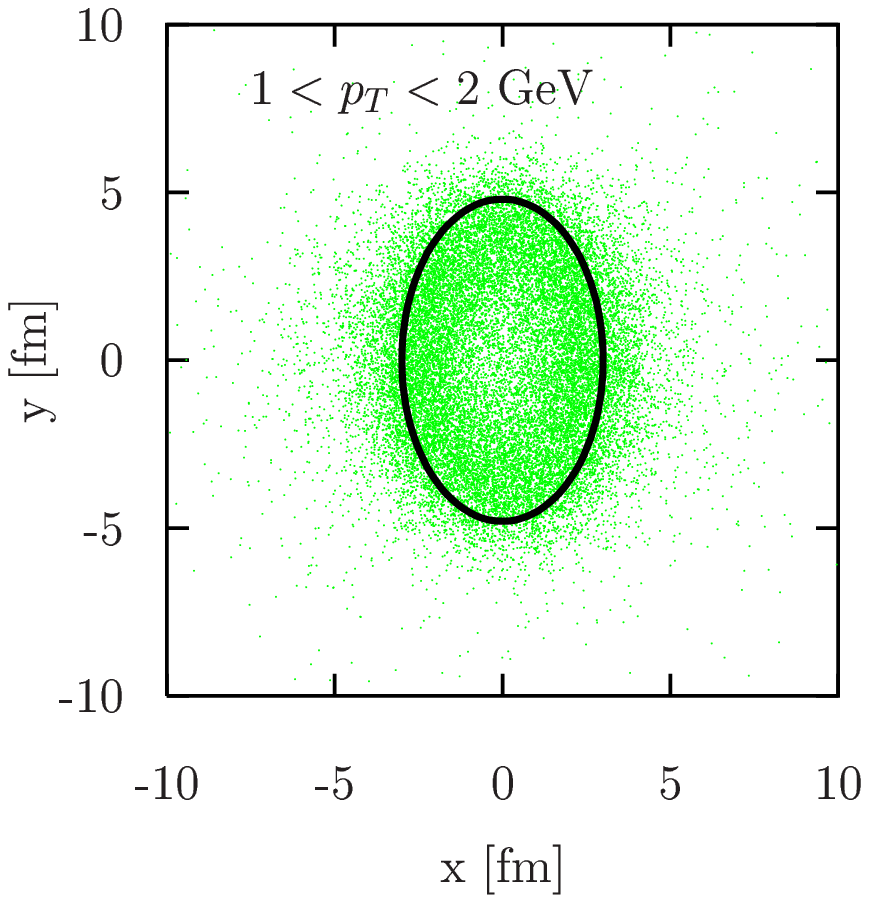,height=1.9in,width=2.in,clip=5,angle=0}
\hskip -0.3cm
\epsfig{file=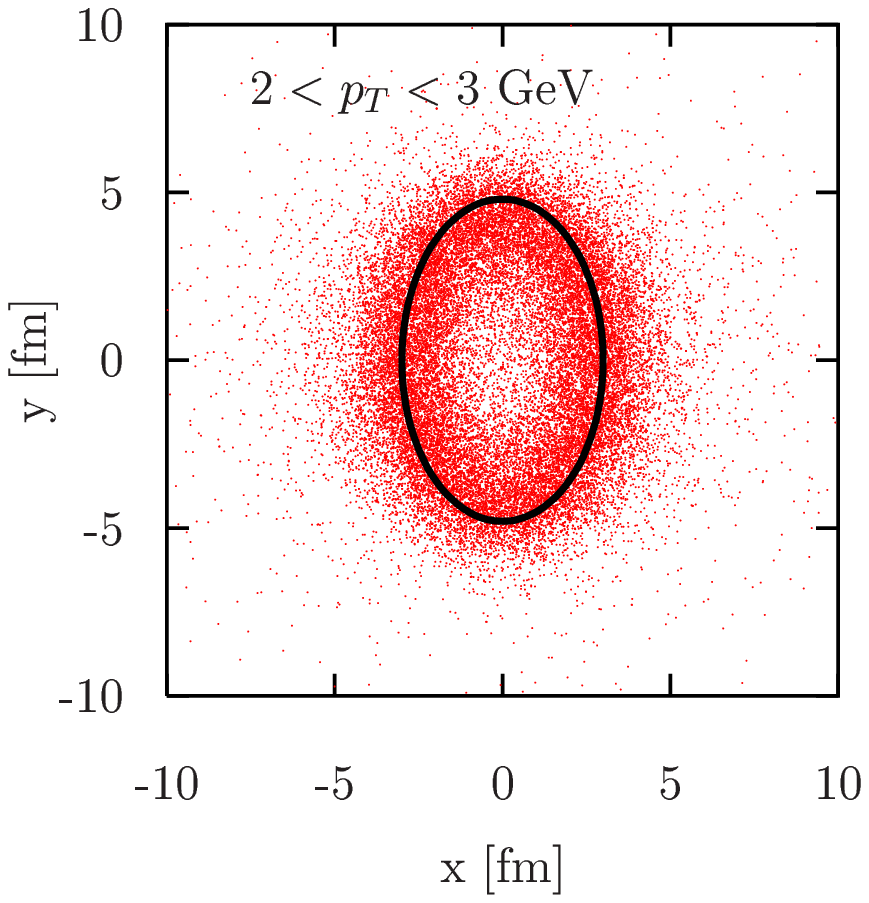,height=1.9in,width=2.in,clip=5,angle=0}
\hskip 1cm
\epsfig{file=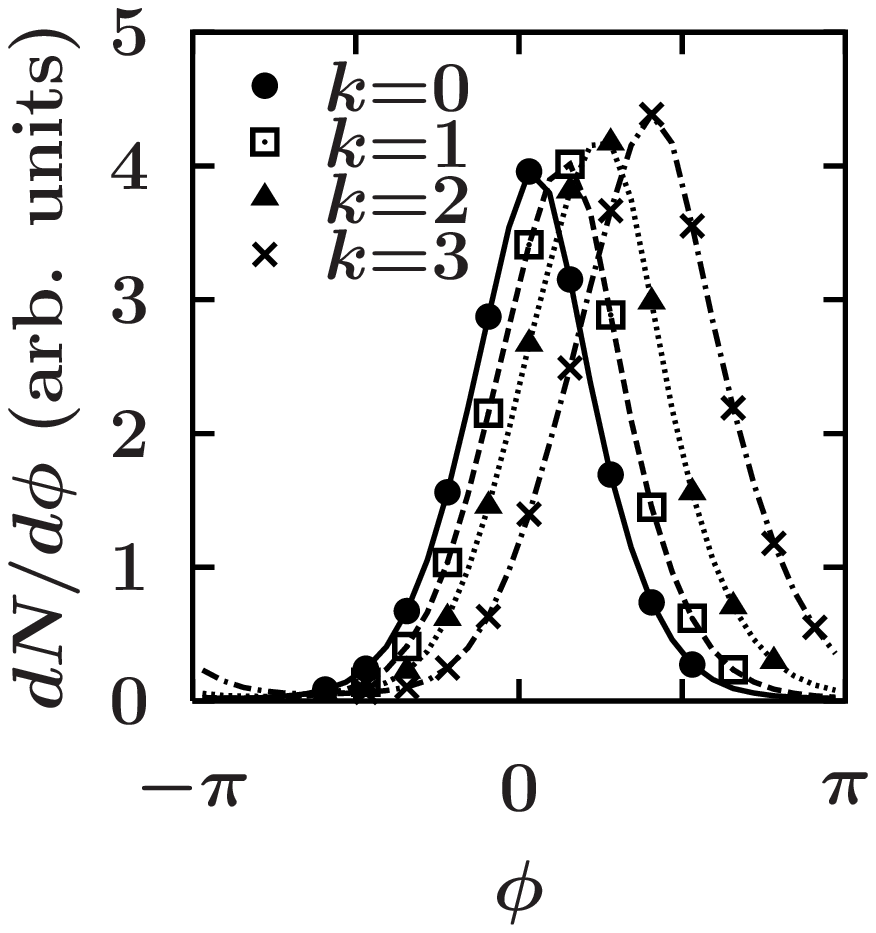,width=1.7in,height=1.8in,angle=0}
\vspace*{-0.2cm}
\hskip 11.5cm a) \hskip 8cm b)
\end{center}
\vspace*{-0.9cm}
\caption{\label{fig:3}
Results from 
MPC\cite{MPC} for $Au+Au$ at $\sqrt{s_{NN}}=200$ GeV with $b=8$ fm.
a) Transverse positions of partons at freezeout as a function
of $p_T$, for final rapidities $|y|<2$ (scatter plot).
The ellipses guide the eye.
b) $\phi$ distributions at midrapidity for $1<p_T<2$ GeV,
averaged over four wedges in the transverse plane 
$\varphi_x \in [k\pi/8, (k+1)\pi/8]$, $k = 0...3$ \cite{v2cvsh}.
}
\end{figure}

Unfortunately, neither of the two cases is realistic.
Dynamical models yield spatially nonuniform local densities {\em and} 
nonuniform flow anisotropies.
Local density variations have been studied 
in connection with the pion interferometry puzzle from
hydrodynamics\cite{KolbHeinzspacetime} and several
transport models\cite{adrianFO,diffuseFO,ziweiFO} 
(Fig. \ref{fig:3}a is an example from covariant parton transport).
Local flow anisotropies are also spatially nonuniform 
as can be seen in Fig. \ref{fig:3}b
that shows (momentum) azimuthal angle $\phi$ distributions 
at midrapidity for $1<p_T<2$ GeV, for $Au+Au$ at top RHIC energy, $b=8$ fm,
averaged over four wedges in the transverse coordinate plane
$\varphi_x \in [k\pi/8, (k+1)\pi/8]$, with $k = 0,1,2,3$ \cite{v2cvsh}.
Clearly, partons emitted from in-plane regions move predominantly in-plane
and, therefore, have positive elliptic flow $v_2 > 0$; while partons
coming from out-of-plane regions move in out-of-plane direction and
have negative $v_2 < 0$.
This is very similar to the hydrodynamic expectation\cite{HeinzHuovinen2source}
based on radially boosted fireballs.

Fig.~\ref{fig:3}b also demonstrates that the condition of small local 
anisotropies (\ref{scalingcond}),
which is necessary for general quark number scaling,
is not satisfied either.
Instead of small harmonic modulations over a uniform background,
the distributions are strongly peaked
because high-$p_T$ particles can only escape from a surface layer of the 
reaction region. 
In this case, local 
anisotropies from coalescence follow a unique power-law scaling~\cite{v2cvsh}
$|v_{had}(p_T)| \simeq |v_{2,q}(p_T/n)|^{1/n}$,
instead of the linear one derived in \cite{coalv2}.

Therefore, quark number scaling of elliptic flow
requires a highly nontrivial, fortuitous 
interplay between variations in local parton
density and nonlinear couplings between large local flow coefficients.
Several classes of phase space distributions have been explored recently
\cite{MSUv2scaling,Texasv2scaling}
and solutions have been found that scale approximately.

One particularly simple example\cite{Texasv2scaling}, 
with large local flow anisotropies, is when $N$ 
partons move along the $\phi = 0$ axis and $(1-a) N$ partons move along 
$\phi = \pi/2$ $(a\ll 1)$. 
If the local volumes for both components are identical,
the hadron yields are $N_{had} \propto V(N/V)^n + V((1-a)N/V)^n$,
and the elliptic flows do scale
$v_2 = [(N/V)^n - ((1-a)N/V)^n]/[(N/V)^n + ((1-a)N/V)^n] \approx n a/2$.
However, it is easy to show that 
scaling crucially depends on the assumption of identical volumes.
Different local volumes, $V$ and $V'$, give hadron yields
$N_{had} \propto V(N/V)^n + V'((1-a)N/V')^n$ and elliptic flows that 
{\em do not scale},
$v_2 = [(V/V')^{n-1} - (1-a)^n]/[(V/V')^{n-1}+(1-a)^n]
\approx [(V/V')^{n-1}-1+na)]/[(V/V')^{n-1} + 1 + na)]$.
This underscores the importance of going beyond simple parameterizations 
and using realistic 
dynamical models to study parton coalescence in heavy-ion
collisions. 

It is important to realize that quark number scaling from coalescence 
necessitates nonequilibrium dynamics\cite{v2cvsh}.
For thermal constituent distributions, 
coalescence reduces to statistical hadronization
and, therefore, momentum anisotropies in that case can only depend on
particle mass, and not quark number. 
This is so even for nonequilibrium 
quark distributions that are of the form $f(x,\vp)=g(p_\mu u^\mu(x),x)$.
Therefore, besides the break-down of 
hydrodynamic behavior, quark number scaling indicates a marked departure
from pure ``flow-like'' space-momentum correlations as well.

\section{Dynamical coalescence approach}
\label{formalism}

One way to study realistic dynamical effects is to combine
nonequilibrium parton transport theory and coalescence~\cite{dyncoal}.
Such an approach requires overcoming the limitations of (\ref{coaleq})
and proper book-keeping to ensure that each parton participates in
one hadronization channel (no multiple counting).

The simple coalescence formula (\ref{coaleq}) relies on the assumption that
interactions between quarks that are not in the same hadron
{\em cease suddenly} on some 3D hypersurface.
However, self-consistent freezeout from transport approaches 
gives diffuse 4D freezeout 
distributions\cite{adrianFO,diffuseFO,ziweiFO}
that cannot be well approximated with a hypersurface.
The coalescence formalism was extended to such a case
 by Gyulassy, Frankel and Remler (GFR)
in \cite{GFR} (for weakly-bound states).
Their result is the same as (\ref{coaleq}),
{\em except} 
that the weight $W$ is evaluated using the {\em freezeout coordinates} 
$(t,\vx)$ of constituents.
When taking $\Delta \vx$, the earlier particle needs to be propagated to the
time of the {\em later} one, e.g.,
$\Delta \vx_{12} = \vx_1 - \vx_2 + (t_2 - t_1) \vv_1$ if $t_1 < t_2$.
The origin of this correction is that a weak bound state 
survives only if none of its constituents have any further interactions.
For baryons, the generalization involves propagation to the {\em latest} 
of the three freezeout times.

To facilitate
proper book-keeping, it is common in transport models\cite{Nagle}
to utilize box Wigner functions
$W=\prod_{i,j} \Theta(x_m-|\Delta \vx_{ij}|)\Theta(p_m-| \Delta \vp_{ij}|)$.
This way (\ref{coaleq}) has a simple interpretation: if $W=1$ 
(and the quantum numbers match) the hadron is formed, 
otherwise it is not ($W=0$).
If several coalescence final states exist
for a given constituent, one is chosen randomly.
Partons that do not find a coalescence partner would propagate ``freely''
and are therefore fragmented independently.

\begin{figure}[h] 
\begin{center}
\epsfig{file=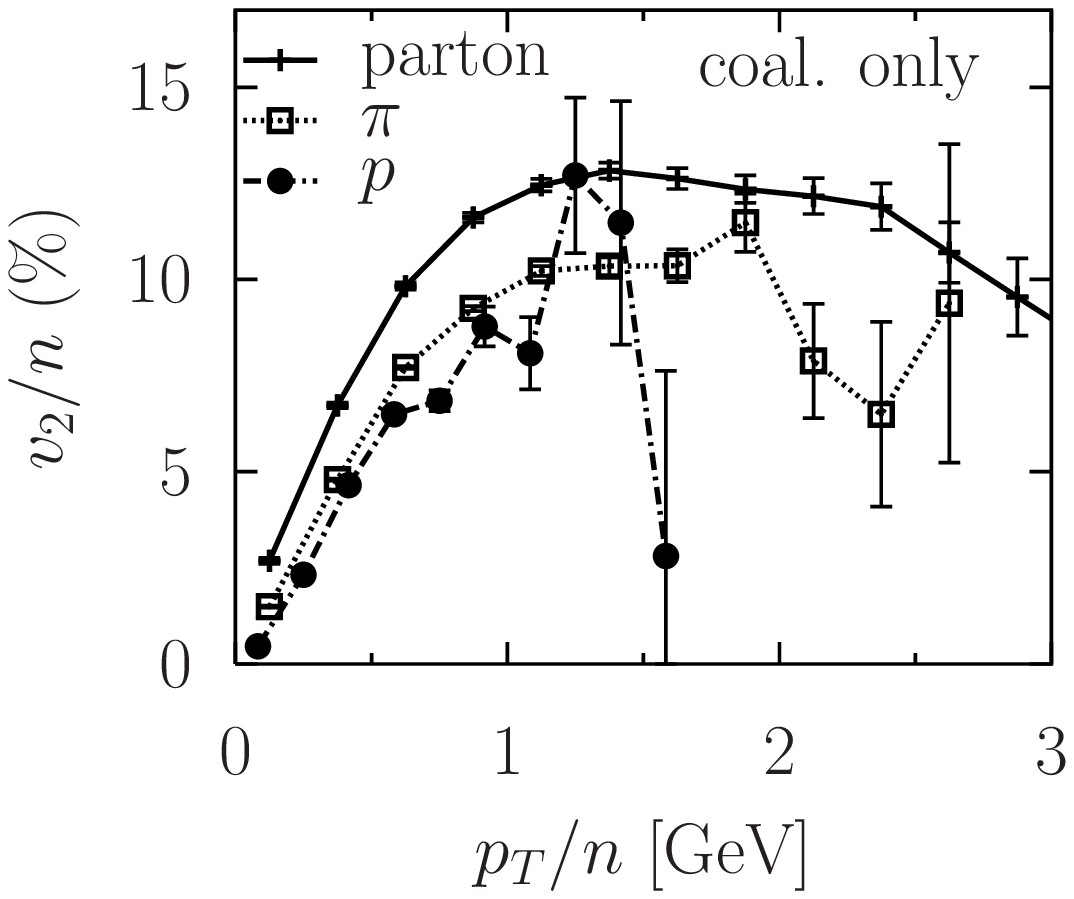,height=1.85in,width=2.45in,clip=5,angle=0}
\epsfig{file=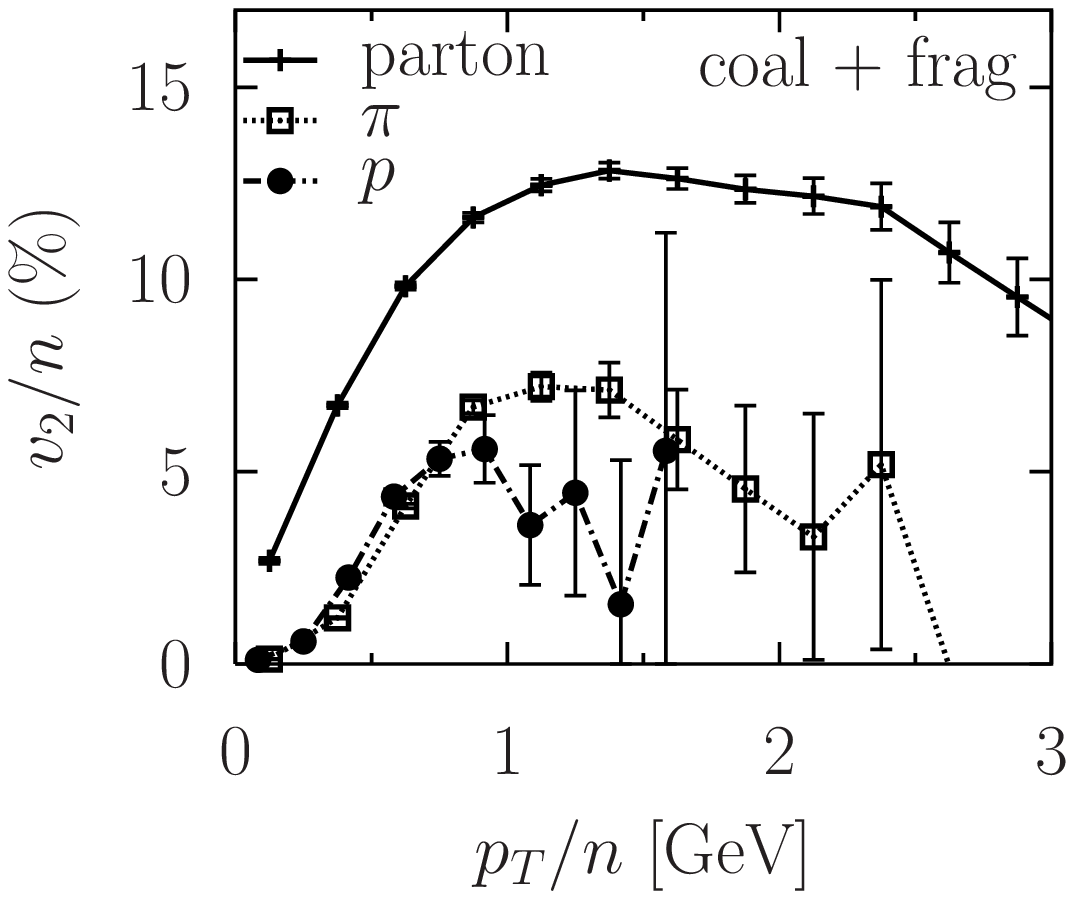,height=1.85in,width=2.45in,clip=5,angle=0}
\end{center}
\vspace*{-0.6cm} 
\caption{\label{fig:4}
Quark number scaled elliptic flow from MPC\cite{MPC} 
as a function of $p_T$ for pions 
(open squares) and protons (filled circles)
in $Au+Au$ at $\sqrt{s}=200A$ GeV at RHIC
with $b=8$ fm and $\sigma_{gg}=10$ mb,
with hadronization via combined coalescence and fragmentation (right),
and for primary hadrons (without decays) from coalescence (left).
The constituent $v_2(p_T)$ is also shown (solid lines). 
Taken from \cite{dyncoal}.
}
\end{figure}

Fig.~\ref{fig:4} shows pion and proton elliptic flow at RHIC
for $Au+Au$ at $\sqrt{s_{NN}}=200$ GeV with $b=8$ fm 
and $\approx 7$ times perturbative opacities ($\sigma_{gg\to gg} = 10$ mb),
from the dynamical coalescence approach\cite{dyncoal} 
based on the covariant transport model MPC\cite{MPC}.
The elliptic flow of {\em direct}
pions and protons from coalescence (left panel) is {\em somewhat smaller}
than the parton flow scaled by constituent quark number.
The difference is due to genuine dynamical effects that are much larger than 
few-percent corrections nonlinear in $v_2$ 
or those due to higher-order flow anisotropies\cite{coalv2}.
Nevertheless, in view of the discussion in the previous Section,
it is quite remarkable that the end result 
is only a modest $\sim 20$\% and $\sim 30$\% 
flow reduction for mesons and baryons.

\begin{figure}[hbpt] 
\begin{center}
\hspace*{-0.5cm}\epsfig{file=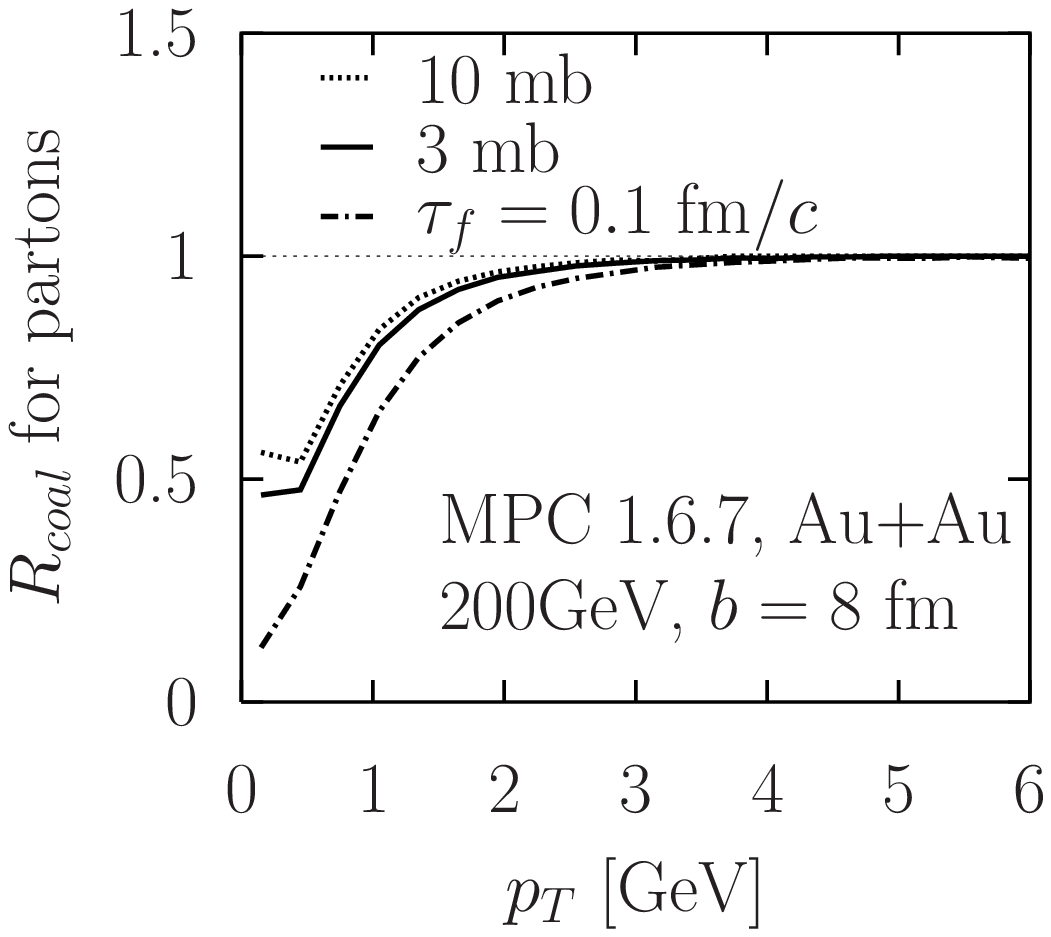,height=1.85in,width=2in,%
clip=5,angle=0}
\hspace*{1.5cm}\epsfig{file=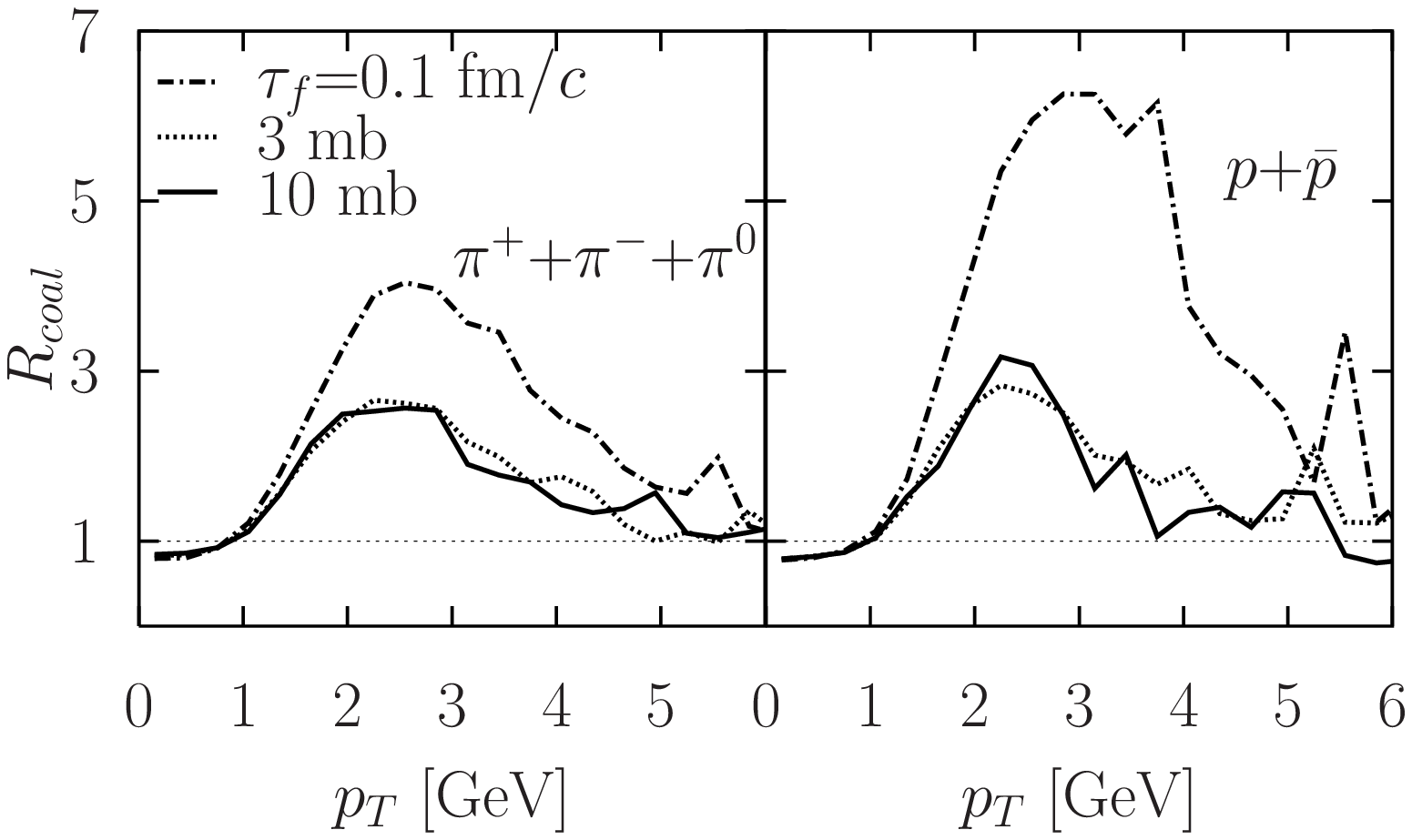,height=1.85in,%
width=2.95in,clip=5,angle=0}
\vspace*{-0.4cm} 
\end{center}
\vspace*{-0.3cm}
\hskip 2.05cm a) \hskip 8.cm b)
\vspace*{-0.5cm} 
\caption{\label{fig:5}
Results \cite{dyncoal} 
from MPC for $Au+Au$ at $\sqrt{s}=200A$ GeV at RHIC with $b=8$ fm 
for $\sigma_{gg} = 3$ mb (dotted) and 10 mb (solid),
or immediate freezeout at $\tau = 0.1$ fm/c (dashed-dotted line).
a) Fraction of partons that fragment independently as a function of $p_T$;
b) pion and proton enhancement from coalescence
as a function of $p_\perp$.
}
\end{figure}

On the other hand, fragmentation contributions sharply
reduce the hadron $v_2$ as shown in the right panel in Fig.~\ref{fig:4}.
Unlike the flow amplification from coalescence\cite{coalv2}, 
fragmentation smears out 
the anisotropy because hadrons from the parton shower are not fully 
collinear (nonzero jet width).
Because only a smaller fraction of parton finds a coalescence partner,
a large fraction fragments, 
including essentially all partons above $p_T > 2.5$ GeV,
as can be seen
in Fig.~\ref{fig:5}a.
Coalescence does become more and more important at low 
$p_T$ but it involves only about one in every three partons.

Despite the large fragmentation component, pion and proton 
$v_2$ still scale approximately {\em relative to each other} -
up to small differences of $\sim 10$\%, perhaps 
as much as 30\%,
which could be tested against high-precision Run-4 data from RHIC 
(better theory accuracy in the $p_T/n > 1-1.5$ GeV region
 is certainly desirable).
However, 
quantitative interpretation of the data becomes more complicated because the
scaled hadron flows {\em underpredict} the real parton $v_2(p_T)$
by factor of 2 or more.

Dynamical effects on the baryon/meson ratio are much more striking
as can be seen in Fig.~\ref{fig:5}b.
Here the yield enhancement from coalescence is characterized by $R_{coal}$, 
the ratio of the final 
spectra with hadronization via combined coalescence and fragmentation to that 
with hadronization via fragmentation only.
Parton coalescence enhances both pion and proton yields,
and hence $R_{AA}$,  by as much as a factor of three 
in the ``coalescence window'' \cite{Voloshincoal,coalv2} 
$1.5< p_T < 4.5$ GeV. The additional hadron yield comes dominantly from
partons with $0.5 < p_T < 2$ GeV (see Fig.~\ref{fig:5}a).
The enhancement, however, is  about the same
for both protons and pions, and hence $p/\pi$ stays close to the value 
in $p+p$ collisions.
Even for a (rather unrealistic) scenario with immediate freezeout
on the formation $\tau = 0.1$fm$/c$ hypersurface, which results in very high 
parton densities, the $p/\pi$ ratio is enhanced
relative to $p+p$ 
only $1.5-1.7$ times, much less than the factor $\approx 3$ seen at RHIC 
at this centrality\cite{PHENIXnoBsupp}.

\section{Conclusions}

Parton coalescence is a promising approach to explain the striking
baryon-meson difference in the $Au+Au$ data at intermediate 
$2 < p_T  < 5$ GeV from RHIC.
Studies based on simple parameterizations are quite successful
but are inconsistent with dynamical models.
A dynamical coalescence approach based on parton transport theory
does confirm that coalescence is an important hadronization channel at RHIC.
However, the dynamics significantly affects elliptic flow scaling 
and baryon/meson ratios.
Until a reliable approach can successfully reproduce the data,
``the jury is still out'' regarding coalescence.

Clearly, further studies are needed.
The $2\to 2$ transport 
considered could be an oversimplification,
or the formalism based on weakly-bound states may not apply well to the 
QCD coalescence process.
Long-range correlations could make coalescence nonlocal 
in phase-space (unlike (\ref{coaleq})), 
or help ``optimally match'' coalescence partners to maximize the yields.
Precise experimental data on deviations from quark number scaling,
flavor dependence of observables, 
and the transition to the pure fragmentation regime
at high $p_T$  will provide additional tests of coalescence 
models and further insights into the dynamics at intermediate $p_T$.

\section*{Acknowledgments}
The hospitality of the RIKEN/BNL Research Center, 
where part of this work was done, is gratefully acknowledged.
This work was supported by DOE grant DE-FG02-01ER41190.

\end{document}